# Bridging the AI divide in sub-Saharan Africa: Challenges and opportunities for inclusivity

Masike Malatji
*Digital Transformation and Innovation*
*Graduate School of Business Leadership (SBL), University of South Africa (UNISA)*
Midrand, Johannesburg, South Africa
malatm1@unisa.ac.za
https://orcid.org/0000-0002-9893-9598

*Abstract*—**The artificial intelligence (AI) digital divide in sub-Saharan Africa (SSA) presents significant disparities in AI access, adoption, and development due to varying levels of infrastructure, education, and policy support. This study investigates the extent of AI readiness among the top SSA countries using the 2024 Government AI Readiness Index, alongside an analysis of AI initiatives to foster inclusivity. A comparative analysis of AI readiness scores highlights disparities across nations, with Mauritius (53.94) and South Africa (52.91) leading, while Zambia (42.58) and Uganda (43.32) lag. Quartile analysis reveals a concentration of AI preparedness among a few nations, suggesting uneven AI development. The study further examines the relationship between AI readiness and economic indicators, identifying instances where AI progress does not strictly correlate with Gross Domestic Product per capita, as seen in Rwanda and Uganda. Using case studies of AI initiatives across SSA, this research contextualises quantitative findings, identifying key strategies contributing to AI inclusivity, including talent development programs, research networks, and policy interventions. The study concludes with recommendations to bridge the AI digital divide, emphasising investments in AI education, localised AI solutions, and cross-country collaborations to accelerate AI adoption in SSA.**

## I. INTRODUCTION

Artificial Intelligence (AI) is emerging as a global innovation and economic growth catalyst. However, AI adoption in sub-Saharan Africa (SSA) remains nascent and unevenly distributed [1]. The region faces an AI digital divide, characterised by disparities in access to AI technologies and knowledge due to underlying gaps in digital infrastructure, education, and supportive policies. While global AI investments promise trillions in economic gains, Africa is projected to reap only minimal benefits under current trajectories [1]. This divide has far-reaching implications: countries with limited AI access risk falling behind in innovation, job creation, and socio-economic development.

Many parts of SSA still struggle with basic connectivity. According to the International Telecommunication Union (ITU) [2], only about one-third of Africans use the Internet, making Africa the least connected region globally. To put this into perspective, in Europe and the Americas, including Latin America (LATAM), 87 and 92 per cent of the population use the Internet [2]. In the Arab States and Asia-Pacific countries, approximately two-thirds of the population (70 and 66 per cent, respectively) do so, which aligns with the global average. By contrast, the average figure for Africa is just 38 per cent [2].

A 2024 study by the Global System for Mobile Communications Association (GSMA) [3], an organisation that represents the interests of mobile network operators worldwide, revealed that by the end of 2023, just 27% (320 million people) of SSA's population was online, leaving a 60% (710 million people) usage gap (people with access to mobile broadband but not using it) and a 13% (160 million people) coverage gap (people with no access to mobile broadband at all). The ITU [2] concurs that the most significant coverage gap is in Africa, where 14% of the continent's population still cannot access a mobile broadband network and, therefore, cannot access the Internet. It should also be noted that urban areas enjoy far better connectivity than rural areas [4], and data costs in Africa are among the highest in the world [2]. For example, in 2023, mobile Internet costs in Africa were 12 times higher than in Europe, a gap that increased to 14 times in 2024. These disparities hinder AI development; without reliable Internet and affordable data, African communities cannot access AI tools or contribute data for AI systems.

The significance of AI inclusivity for Africa is profound – inclusive AI adoption could drive improvements in agriculture, health, education, and government services, enabling leapfrog development and more equitable growth [5]. Yet the research problem is that AI resources and knowledge in SSA are unevenly distributed. A few countries (such as South Africa, Kenya, Nigeria, Rwanda, and Mauritius) host the majority of AI initiatives, technology hubs, and expertise, while many others lag. This imbalance threatens to exacerbate inequalities in innovation capacity, employment opportunities, and social services delivery across the continent. If unaddressed, the AI divide could reinforce existing socio-economic divides and leave large segments of the African population excluded from the benefits of the digital revolution [5].

This paper seeks to investigate the extent and impact of the AI digital divide in SSA and to identify strategies for fostering

greater AI inclusivity. Specifically, I ask the following research questions (RQs):

- **RQ1**: What is the current state of AI access and usage across SSA, and which regions or countries lead or lag in AI adoption readiness?
- **RQ2**: What are the main barriers that limit AI inclusivity in the region?
- **RQ3**: What strategies and initiatives can effectively bridge the AI digital divide and promote inclusive AI development in SSA?

The rest of this paper is structured as follows: Section II reviews the existing knowledge on AI digital divide and digital inclusion, AI infrastructure and connectivity disparities in SSA, AI skills gap and educational disparities, and policy and governance challenges for AI inclusivity. In Section III, I map out the methods adopted for answering the three RQs and present the findings and discussions in Section IV. I conclude the paper in Section V, where future research and paper expansion avenues are also proposed.

## II. RELATED WORKS

### A. Defining the AI digital divide and digital inclusion theory

The concept of the digital divide traditionally refers to the gap between those with and without access to information and communication technologies (ICT) [6]. In the context of AI, this has evolved into an AI digital divide – the disparity between individuals, organisations, or nations with access to AI tools, data, and expertise and those without [7]. The AI digital divide builds on earlier digital inclusion frameworks, which emphasise physical access to technology and the skills and meaningful usage required to benefit from it [7], [8]. In the Fourth Industrial Revolution (4IR), this AI divide is widening quickly as AI advancements accelerate in developed regions. At the same time, many developing countries still lack the basic digital infrastructure and skills of the previous revolution.

Theoretical perspectives on digital inclusion suggest a multi-dimensional challenge: beyond basic connectivity (often termed the "first-level" digital divide), there are second-level divides in digital skills, and third-level divides into outcomes or benefits gained [7]. Applied to AI, this means that even as the Internet and devices become available, a gap may remain in the ability to develop or use AI (due to a lack of skills or localised content), and consequently, a gap in deriving economic or social benefits from AI. Inclusivity in AI, or the use of AI for local needs, thus requires addressing the following: (i) access, (ii) capability, and (iii) empowerment.

### B. AI infrastructure and connectivity disparities in SSA

Broadband connectivity and computing infrastructure are foundational for AI adoption [9]. SSA faces significant deficits on these fronts. As mentioned in the introduction section, Internet penetration in Africa is the lowest of any world region defined by the ITU [2]. Only about 38% of African individuals were online as of 2024, compared to a global average of 67% [2]. Fixed broadband is exceptionally scarce (just ~3% subscription rate in Africa vs 15% globally), and while mobile networks have expanded, a large usage gap persists where coverage exists [10]. However, people remain offline. By the end of 2023, only 27% of the population was online, while 60% had network coverage but were not using the Internet, often due to issues like cost, skills, or lack of relevant content [2], [3], [7]. The remaining 13% had no coverage at all. This shows that infrastructure availability, affordability, and adoption are all challenges in SSA.

Thus, high costs and limited infrastructure quality undermine AI readiness. Furthermore, unreliable electricity and sparse data centre infrastructure can further constrain AI development. For instance, almost 600 million Africans lack electricity, representing approximately 60 per cent of the continent's total [11]. Even where consumers are connected to the electric grid, electricity supply tends to be unreliable and plagued by frequent outages. Moreover, African firms report an average of 56 days without electricity annually [11]. On the ICT infrastructure side, local cloud or high-performance computing resources are limited. These infrastructure gaps mean AI researchers and companies in Africa often struggle with access to computing power and must rely on external cloud services, raising costs and latency.

Connectivity disparities also manifest within countries. Urban areas have broadband and fibre connections, technology hubs, and better access to data centres, whereas rural communities often lack even basic 3G/4G coverage [12], [13]. Therefore, this urban-rural divide in Internet access translates into an AI divide, as urban institutions and firms can leverage AI tools that are practically out of reach in rural settings. Furthermore, there are demographic divides: for example, Africa's digital gender gap has been widening – women are less likely to have Internet access and thus are underrepresented in the digital economy [14]. Such disparities mean that AI initiatives in Africa may currently benefit a relatively small urban elite segment, risking the exclusion of rural populations, women, and marginalised groups.

### C. AI skills gap and educational disparities

A critical component of the AI digital divide is the skills gap. Many African universities and schools have yet to integrate modern AI, data science or 4IR training into their curricula [15], [16]. As a result, there is a shortage of AI skills to accelerate the development of AI in Africa [17], and those with advanced skills are concentrated in a few institutions or often pursue opportunities abroad (brain drain). The World Bank estimates that by 2030, Africa will require 230 million digital jobs, most of which will demand a workforce equipped with intermediate to advanced digital skills. However, current educational pipelines are not keeping pace with this demand.

Furthermore, disparities in AI education are evident across countries. Only a handful of universities in SSA offer specialised programs in AI or machine learning (ML), and these are often found in countries like South Africa, Nigeria, Kenya, or Egypt (North Africa) with relatively stronger tertiary education systems. Many smaller or lower-income countries have limited science, technology, engineering, and mathematics (STEM) education and research capacity. This leads to an uneven distribution of AI talent. For instance, hubs like Nairobi in Kenya, Lagos in Nigeria, and Johannesburg in South Africa

[18] produce and attract most of Africa's AI researchers and engineers, while other regions have few experts. Moreover, within countries, access to quality computer science education is typically better in urban centres than in rural areas. The result is a pronounced skills divide: a minority of Africans have the opportunity to acquire AI literacy and expertise, whereas the majority lack even essential AI awareness.

Efforts are underway to bridge the skills gap, but challenges remain. Initiatives like Data Science Africa, a non-governmental organisation established in Kenya that promotes the affordability, broader deployment ability, and applicability of AI technologies in Africa [19], have emerged as grassroots responses, providing hands-on training in ML to students and researchers who might otherwise lack exposure. Data Science Africa's workshops and summer schools, since 2015, have trained participants across eight countries (including Kenya, Nigeria, Ghana, and Rwanda) precisely to compensate for limited university offerings during that decade. Similarly, the annual Deep Learning Indaba fosters a pan-African AI community and mentoring network to build local capacity. For example, Deep Learning Indaba helps to cultivate a community of AI researchers [20], technology startups and AI education enthusiasts [21]. These efforts highlight the need for systemic educational reforms – incorporating AI and data science into secondary and tertiary curricula, investing in faculty development, and creating incentives for skilled professionals to remain in or return to Africa's academia and industry.

### D. Policy and governance challenges for AI inclusivity

Policy frameworks and governance significantly influence AI adoption. Digital policy development has not kept pace with technological change in many African nations. Until recently, only one country in SSA – Mauritius – had a dedicated national AI strategy launched in 2018 [14]. This is slowly changing: Rwanda, Senegal, and Benin each published national AI strategies in 2023, becoming among the first in mainland Africa. Ethiopia, South Africa, Nigeria, and Zambia released their AI strategies in 2024 [22]. These strategies mark important steps toward official recognition of AI's importance. However, most countries still lack clear AI policies or guidelines, and even where strategies exist, implementation and funding remain uncertain. The African Union (AU) has acknowledged the need for a coordinated approach at the continental level. As a result, the AU released its continental AI strategy during its 45th Ordinary Session in Accra, Ghana, on July 18-19, 2024, which aims to harness AI for Africa's development and prosperity (Oxford Insights, 2024).

Pan-African policy bodies and international partners like the ITU and the United Nations Educational, Scientific and Cultural Organisation (UNESCO) advocate for frameworks that ensure AI is developed ethically and inclusively in Africa [14]. Despite this, coherent governance is hindered by limited institutional capacity and cross-country coordination issues. Another governance challenge is ensuring that AI deployment in Africa aligns with local needs and values. Much of the AI technology used in Africa is imported or developed by foreign companies [14].

This raises concerns that outside solutions may not fully address local priorities or could even entrench biases and inequalities, for example, AI systems that do not recognise African languages or that reflect Western biases. Without local participation in AI design, Africans risk being passive consumers of AI with little influence over its impacts. Additionally, regulatory gaps (such as data protection, cybersecurity, and AI ethics) could stifle innovation or expose societies to harm [23], [24]. For example, few countries have comprehensive data privacy laws, yet data governance is crucial for building public trust in AI.

In summary, the literature indicates that bridging the AI digital divide in Africa requires simultaneous progress on multiple fronts: (i) upgrading infrastructure, (ii) building human capital, and (iii) crafting enabling policies. The following section will detail the research approach to examining these issues and present findings on how these factors interplay across SSA.

### III. METHODS

A mixed-methods research design was adopted, combining quantitative data analysis with qualitative case studies. The study's approach is outlined in Fig. 1.

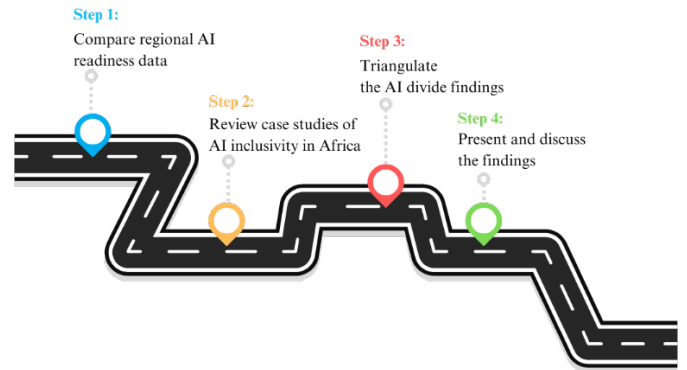

Fig. 1. Research approach roadmap

As shown in Fig. 1, a comparative regional analysis study was conducted. I gathered secondary data from Oxford Insights [22] to compare AI readiness across sub-Saharan African countries. Specifically, I performed a comparative analysis of the difference between the top and bottom countries, a quartile analysis, and a ranking versus the regional development index. The regional development index (Gross Domestic Product (GDP) per capita) data was obtained from Trading Economics (2023) and is measured in the United States of America's Dollar (USD) currency. This provided a macro-level view of the AI access gap, highlighting which countries in the region are relatively ahead and which are lagging.

As the second step, qualitative case studies were conducted on successful (or promising) initiatives to bridge the AI divide in SSA. These are shown in Table I.

TABLE I. CASE STUDIES OF AI INCLUSIVITY INITIATIVES

| Initiative Name | Country | City | URL |
|---|---|---|---|
| **Deep Learning Indaba** | Pan-African | Various | https://deeplearningindaba.com/ |

| Name | Country | City | URL |
|---|---|---|---|
| **Data Science Africa** | Pan-African | Various | http://www.datascienceafrica.org/ |
| **AI4D Africa** | Pan-African | Various | https://ai4d.ai/ |
| **Google AI Research Center** | Ghana | Accra | https://ai.google/research/locations/accra/ |
| **Microsoft Africa Development Center** | Kenya, Nigeria | Nairobi, Lagos | https://www.microsoft.com/en-africa/adc |
| **African Masters of Machine Intelligence (AMMI)** | Senegal | Dakar | https://aimsammi.org/ |
| **Zindi Africa** | Pan-African | Various | https://zindi.africa/ |
| **Masakhane** | Pan-African | Various | https://www.masakhane.io/ |
| **AI Expo Africa** | South Africa | Cape Town | https://aiexpoafrica.com/ |
| **AI Media Group** | South Africa | Johannesburg | https://aimediagroup.co.za/ |
| **Artificial Intelligence for Development (AI4Dev)** | Uganda | Kampala | https://ai4dev.ug/ |
| **Rwanda Artificial Intelligence Research Group (RAIRG)** | Rwanda | Kigali | https://www.rairg.rw/ |
| **Ghana Natural Language Processing** | Ghana | Accra | https://ghananlp.org/ |
| **Nigerian Data Science and Artificial Intelligence Summit (DSAIS)** | Nigeria | Lagos | https://www.datasciencenigeria.org/dsais/ |
| **AI Kenya** | Kenya | Nairobi | https://www.aikenya.co.ke/ |
| **South African Centre for Artificial Intelligence Research (CAIR)** | South Africa | Various | https://www.cair.org.za/ |
| **IndabaX** | Various | Various | https://indabax.africa/ |
| **AI Saturdays Lagos** | Nigeria | Lagos | https://www.aisaturdayslagos.com/ |
| **Machine Learning Zimbabwe** | Zimbabwe | Harare | https://mlzimbabwe.com/ |
| **AI Expo Botswana** | Botswana | Gaborone | https://www.aiexpoafrica.com/ |
| **Nairobi Women in Machine Learning & Data Science** | Kenya | Nairobi | https://www.meetup.com/Nairobi-Women-in-Machine-Learning-Data-Science/ |
| **Data Science Nigeria** | Nigeria | Lagos | https://www.datasciencenigeria.org/ |
| **Ghana AI Meetup** | Ghana | Accra | https://www.meetup.com/Ghana-AI/ |
| **Tanzania Data Lab (dLab)** | Tanzania | Dar es Salaam | https://dlab.or.tz/ |
| **Ethiopia Artificial Intelligence Research Center** | Ethiopia | Addis Ababa | http://www.eairc.gov.et/ |
| **AI Nigeria** | Nigeria | Lagos | https://ainigeria.ng/ |
| **Kigali Artificial Intelligence Research Institute (Kigali AI)** | Rwanda | Kigali | https://kigaliai.rw/ |
| **AI Robotics Nigeria** | Nigeria | Lagos | https://airobotics.ng/ |
| **Machine Intelligence Institute of Africa (MIIA)** | South Africa | Cape Town | https://miia.io/ |
| **AI Expo East Africa** | Kenya | Nairobi | https://www.aiexpoafrica.com/ |
| **Nigerian Association of Computational Linguistics (NACL)** | Nigeria | Abuja | https://nacl.org.ng/ |
| **AI Research Lab Makerere University** | Uganda | Kampala | https://air.ug/ |
| **Botswana AI Society** | Botswana | Gaborone | https://www.botswanaaisociety.org/ |
| **AI Senegal** | Senegal | Dakar | https://aisenegal.org/ |
| **AI South Africa** | South Africa | Johannesburg | https://www.aisouthafrica.com/ |
| **AI Mali** | Mali | Bamako | https://aimali.org/ |
| **AI Ghana** | Ghana | Accra | https://aighana.org/ |
| **AI Liberia** | Liberia | Monrovia | https://ailiberia.org/ |
| **AI Côte d'Ivoire** | Côte d'Ivoire | Abidjan | https://aicotedivoire.org/ |
| **AI Burkina Faso** | Burkina Faso | Ouagadougou | https://aiburkinafaso.org/ |
| **AI Niger** | Niger | Niamey | [https://ainiger.org/](https://ain |

Thirty-five Like Deep Learning Indaba and Data Science Africa, the initiatives foster inclusive AI development in SSA by enhancing education, infrastructure, policy frameworks, and other aspects. For each case, I documented the context, goals, and outcomes to distil lessons on what works in fostering inclusive AI. As the third step of the study, I triangulated findings from the two methods to increase credibility and validity. Quantitative data established the broad patterns of the AI divide in SSA, while case studies helped explain the underlying causes and contextual nuances. This approach allowed me to identify where and how the gaps are, why they persist, and what interventions might be effective. The fourth and final step of the methodology depicted in Fig. 1 follows in the next section.

IV. FINDINGS AND DISCUSSION

*A. AI readiness index scores*

The Oxford Insights' Government AI Readiness Index report looks at 40 indicators across three pillars: Government, Technology Sector, and Data and Infrastructure. This index measures governmental AI implementation readiness in public services, providing a critical tool for evidence-based policymaking to maximise AI's potential for citizen benefit. The regional analysis section of the 2024 report shows that the SSA has the least prepared AI infrastructure globally [22], mainly due to the region's high cost, low quality, and low availability of Internet connectivity. The top ten SSA countries and their respective AI readiness scores are listed in Table II. These were obtained from the Oxford Insights' [22] Government AI Readiness Index report.

TABLE II. TOP TEN SSA COUNTRIES AND THEIR RESPECTIVE AI READINESS SCORES

| Rank | Country | AI readiness score |
|---|---|---|
| 1 | Mauritius | 53.94 |
| 2 | South Africa | 52.91 |
| 3 | Rwanda | 51.25 |
| 4 | Senegal | 46.11 |
| 5 | Benin | 45.62 |
| 6 | Ghana | 44.98 |
| 7 | Kenya | 44.79 |
| 8 | Seychelles | 44.77 |
| 9 | Uganda | 43.32 |
| 10 | Zambia | 42.58 |

Analysing the 2024 Government AI Readiness Index for the top ten SSA countries provides insights into disparities and correlations with economic indicators. Starting with the difference between top and bottom countries, the AI readiness scores range from 53.94 (Mauritius) to 42.58 (Zambia), indicating a disparity of 11.36 points. Secondly, I looked at the quartile analysis to help understand whether AI readiness is evenly distributed or concentrated among a few leading countries. I divided the ten countries into quartiles as follows:

- Q1 (Top 25%): Mauritius (53.94), South Africa (52.91), Rwanda (51.25)
- Q2: Senegal (46.11), Benin (45.62)
- Q3: Ghana (44.98), Kenya (44.79), Seychelles (44.77)
- Q4 (Bottom 25%): Uganda (43.32), Zambia (42.58)

The scores are relatively close, with a slight concentration in the top quartile. This answers RQ1 on the countries leading or lagging in AI adoption readiness. Lastly, I looked at the ranking versus the regional development index to assess correlation, as shown in Table III.

TABLE III. COMPARISON OF AI READINESS SCORES WITH GDP PER CAPITA

| Country | AI readiness score | GDP per capita |
|---|---|---|
| Mauritius | 53.94 | 11,319 |
| South Africa | 52.91 | 5,747 |
| Rwanda | 51.25 | 1,004 |
| Senegal | 46.11 | 1,463 |
| Benin | 45.62 | 1,300 |
| Ghana | 44.98 | 2,087 |
| Kenya | 44.79 | 1,808 |
| Seychelles | 44.77 | 16,715 |
| Uganda | 43.32 | 956 |
| Zambia | 42.58 | 1,331 |

Table III shows that Mauritius and Seychelles have high GDP per capita and AI readiness. On the other hand, Rwanda and Uganda have low GDP per capita but relatively high AI readiness, suggesting effective AI strategies despite economic challenges.

This analysis highlights the complex relationship between economic development and AI readiness in SSA. Discussing these rankings in general, Table II showed each country's preparedness to implement AI in public services, considering factors such as government strategy, technology infrastructure, and data availability [22]. Table II further showed that countries such as Mauritius, South Africa, and Rwanda stand out as front-runners, with clear momentum in strengthening their AI ecosystems [22].

Mauritius excels particularly in government support for AI, while South Africa scores highest in technology infrastructure and is the only SSA nation scoring above the global average (42.55) in that area. Furthermore, these countries tend to have better Internet and power infrastructure, more robust higher education in STEM, and proactive government or private sector initiatives in AI. They also host numerous technology hubs – for instance, South Africa, Nigeria, Kenya, and Egypt, forming an "innovation quadrangle" with over 80 active technology hubs collectively, indicating concentrated digital innovation ecosystems. However, the report shows that despite their highest AI readiness rank in SSA, these countries still have digital infrastructure scores well below the global average.

By contrast, many African countries show very low AI readiness. Many nations (not shown in Table II, as SSA has roughly 49 countries) in the bottom ranks of the global index are in SSA, reflecting limited data infrastructure and human capital for AI. Countries such as Eritrea, Burundi, South Sudan, and others affected by conflict or low development have almost negligible AI activity. Even larger economies like the Democratic Republic of Congo or Ethiopia, despite population size, lag in AI due to low Internet penetration and investment. However, as mentioned in the literature, Ethiopia is one of the four SSA countries that have released their AI strategies in 2024. Nonetheless, the regional average for SSA on the AI readiness index is the lowest among the nine world regions. This highlights that even the frontrunners in Africa fall far behind Europe, North America, and Asia-Pacific countries.

The literature mentions that the main barriers to AI inclusivity can be divided into three levels:

- Level 1 AI divide: lack of basic connectivity (e.g., Inadequate or poor infrastructure, high data costs, lack of access to AI tools and devices, etc.)

- Level 2 AI divide: lack of skills (e.g., inadequate digital and AI literacy, shortage of professionals with AI skills, limited institutional capacity, limited cross-country coordination, imported AI technology which seldom aligns with local needs and values, for example, localised content)
- Level 3 AI divide: Policies and governance (e.g., regulatory gaps in data protection, cybersecurity and AI ethics, lack of clear AI policies or guidelines)

In summary, the main barriers to AI inclusivity are: (i) inadequate infrastructure, (ii) skill and education gaps, (iii) policy and regulatory gaps, (iv) funding and investment constraints, and (v) language and local content. This answers RQ2 on the main barriers that limit AI inclusivity in SSA.

### B. AI inclusivity development case studies

Table IV analyses selected AI initiatives across SSA, detailing their context, goals, and outcomes. This analysis evaluates how these initiatives contribute to inclusive AI development by enhancing education, infrastructure, and policy frameworks. These were selected for their continental-wide orientation as opposed to country-specific initiatives. However, it must be emphasised that SSA has more AI/ML and data science initiatives than shown in Table IV.

TABLE IV. CONTEXT, GOAL(S) AND OUTCOME OF SELECTED AI INITIATIVES

| Initiative name | Context | Goal | Outcome |
|---|---|---|---|
| Deep Learning Indaba | Established to strengthen African ML and AI communities by connecting researchers and practitioners. | To enhance knowledge sharing, capacity building, and mentorship in AI across Africa. | Conducted annual gatherings, workshops, and mentorship programs, fostering a robust AI community. |
| Data Science Africa | Focusing on capacity building in data science and AI/ML through practical training and workshops. | To promote data science research and application for addressing local challenges in Africa. | Organised hands-on workshops and conferences, leading to increased local expertise and research collaborations. |
| AI4D Africa | A network aiming to support AI innovations that address development challenges specific to Africa. | To foster locally relevant AI research and applications for sustainable development. | Funded projects tackling issues like agriculture, health, and language processing, contributing to localised AI solutions. |
| Zindi Africa | A data science competition platform connecting African data scientists to solve real-world problems. | To provide opportunities for data scientists to develop solutions for African challenges and gain recognition. | Hosted numerous competitions, resulting in innovative solutions and skill development among participants. |
| Masakhane | A research community focused on natural language processing (NLP) for African languages. | To develop NLP technologies that cater to African languages, promoting linguistic diversity in AI. | Produced open-source datasets and models for multiple African languages, enhancing digital inclusivity. |

As Table IV indicates, there are ongoing efforts to bridge the AI divide. The Deep Learning Indaba initiative was launched in 2017 as a pan-African annual gathering of AI/ML researchers and students to address the severe underrepresentation of African voices in the AI field [20], [21]. It provides training, networking, and a prestigious platform for African AI/ML research (through its "Research in Africa" showcase). Crucially, Indaba has spawned IndabaX – locally organised workshops in dozens of countries (47 countries held IndabaX events in 2024) to democratise AI knowledge at the grassroots [25]. This decentralisation means even countries with no formal AI institutes can have an annual mini-conference and training session, empowering students and developers. Likewise, Data Science Africa conducts open summer schools and hackathons that have trained hundreds on practical ML, explicitly targeting the gap in university curricula [19]. These programs, along with meetups and online communities, are slowly expanding the base of AI-literate Africans and seeding research relevant to African problems (e.g., crop disease detection or local language translation).

International support programs also contribute. For example, AI for Development (AI4D) Africa and Deep Learning Indaba have received backing from international technology companies and global charities, bringing resources and mentorship. There are also emerging government-led interventions: countries like Kenya and Ghana have set up innovation hubs with government support to incubate startups in AI and data analytics. The African Development Bank's investments in digital centres of excellence and the World Bank's funding for African digital transformation (such as the Digital Economy for Africa initiative) are channelling funds into expanding broadband and training digital skills, indirectly supporting AI readiness.

The discussion above, which answered RQ1 and RQ2, indicates that while challenges are daunting, they are increasingly recognised, and a combination of bottom-up and top-down efforts is underway. To answer RQ3 on the strategies and initiatives that can effectively bridge the AI divide and promote inclusive AI development in SSA, the following section builds on insights from the discussions above.

### C. Bridging the AI divide

The uneven AI adoption readiness landscape means that AI innovation (research labs, startups, AI conferences) is heavily concentrated in a few locales like Nairobi, Lagos, Dakar, or Johannesburg and Cape Town, with a sparse presence elsewhere. Thus, bridging the AI digital divide in SSA will require coordinated action across policy, education, and infrastructure domains. Based on the literature review and my findings, I propose five key strategies:

- Strengthen digital infrastructure as a foundation
- Develop inclusive AI policies and strategies
- Build AI capacity through education and upskilling
- Foster local AI ecosystems and research
- Ensure inclusive governance and ethics

Starting with strengthening digital infrastructure as a foundation, SSA governments and regional bodies should prioritise investments in connectivity and computing infrastructure. This includes expanding broadband Internet access to underserved areas (through public-private partnerships in fibre optics, satellite broadband, or community networks) and improving the affordability of data (such as reducing taxes on Internet services or supporting free public Wi-Fi zones). Additionally, AI-specific infrastructure needs attention: establishing data centres and cloud computing facilities within SSA can significantly lower local AI development barriers. For example, investment in high-performance computing clusters or university Graphics Processing Unit (GPU) farms would enable researchers to experiment with advanced AI models without needing overseas resources. International initiatives could support this, e.g., partnerships with global cloud providers to create regional AI innovation centres that offer computing credits to startups and researchers.

In developing inclusive AI policies and strategies, every country in SSA should formulate a national AI strategy or include AI in its digital development plans (see Table 2). These policies must emphasise inclusivity, ensuring AI benefits reach urban and rural, men and women, rich and poor. Key policy measures should involve:

- Setting up dedicated AI task forces or agencies to coordinate AI initiatives across ministries (education, ICT, industry, etc.).
- Implementing regulations encouraging innovation while protecting citizens, such as clear data protection laws (aligned with the AU's Malabo Convention and international best practices) to build trust for data sharing and sandboxes for AI startups to test new solutions safely.
- Creating incentive programs (e.g., grants, tax breaks) for businesses and social enterprises working on AI solutions that address local challenges, such as AI in agriculture or healthcare for low-income communities.
- Adopting open data policies so that government datasets, for example, in agriculture, health, education, etc., are accessible for building AI applications – this can stimulate local innovation and ensure AI models are trained on African-relevant data.
- Engaging in international and regional cooperation, like contributing to the AU's AI blueprint, sharing resources, and aligning with ethical guidelines. Regional centres of excellence could be established under the AU or regional economic communities to support smaller countries.

To close the skills gap, a pipeline of AI talent must be nurtured to build AI capacity through education and upskilling SSA countries. This starts with integrating digital literacy and coding at the primary and secondary school levels so that children grow up comfortable with the concepts behind AI. Secondary schools can introduce basic AI and robotics clubs to spark interest. At the tertiary level, universities should be supported, via funding and partnerships, to update curricula and launch programs in AI, data science, and related fields. This might involve twinning programs with foreign universities, industry-sponsored labs, or training trainers to build teaching capacity rapidly. Scholarships and fellowships dedicated to AI studies can encourage more students to specialise in this area. Moreover, massive open online courses (MOOCs) and certification programs should be promoted beyond formal education to upskill the current workforce, ensuring that even graduates can acquire new AI skills. Initiatives like the ITU's AI Skills Coalition, launched in 2025 to provide open training in machine learning and AI for all, are valuable and should be localised into African languages and contexts. Countries can also leverage technology hubs and innovation labs as informal training grounds, supporting hackathons, boot camps, and mentorship networks that reach youth outside the academic system.

Creating an environment where African AI innovations can flourish is key to inclusivity. This can foster local AI ecosystems and research. It entails establishing more technology hubs, incubators, and research institutes focused on AI across the continent. Governments and investors can provide seed funding for AI startups working on local issues, for example, fintech solutions for financial inclusion, language translation tools, or AI for climate-smart agriculture. Collaboration between universities and industry should be incentivised to ensure research and development (R&D) translates into applied solutions. Also, South-South collaborations within Africa can help share expertise. For example, a well-equipped AI lab in one country could host trainees or joint projects with institutions in neighbouring countries that lack such facilities. Continent-wide workshops like Deep Learning Indaba and networks such as Data Science Africa and Masakhane for NLP (see Table 4) should continue to be supported and expanded as they build critical mass and community among practitioners. By nurturing local innovation, SSA can produce AI applications tailored to local languages, cultural contexts, and pressing development needs rather than relying solely on imported AI products.

As AI adoption increases, it is crucial to do so in an equitable manner that respects human rights. Ensuring inclusive governance and ethics of AI implementations is therefore paramount. Governments should incorporate ethical guidelines in line with UNESCO's recommendation on AI ethics and other frameworks into their AI strategies, focusing on transparency, accountability, and fairness. This includes monitoring for biases in AI systems that could disadvantage any group and mandating stakeholder engagement, including civil society, in AI policy-making. Another strategy is to launch public awareness campaigns about AI, demystifying AI for the general population and encouraging participatory discussions on how AI should be used in society. An informed populace can better advocate for their needs and hold institutions accountable. For inclusivity, special attention should be paid to gender inclusion in AI, supporting women-in-STEM initiatives, such as accessibility (ensuring AI tools are accessible to those with disabilities and available in local languages). In summary, inclusive governance means not only expanding AI's reach but doing so in a way that protects against new inequalities.

Implementing these key strategies will require concerted effort and significant resources, but the opportunities are vast. With supportive policies and investments, AI can be a leapfrogging tool for SSA – improving crop yields through precision farming, expanding healthcare access via AI diagnostics and telemedicine, and creating new jobs in the ICT sector that were previously unimaginable. Notably, a focus on inclusivity ensures that these benefits are widely shared and do not merely accrue to already advantaged groups.

## V. Conclusion

This paper has examined the challenges and opportunities in bridging the AI divide in SSA. AI adoption in the region is characterised by stark disparities: a minority of countries and urban centres are making strides in AI, while many others remain on the sidelines of this technological revolution. Key barriers include inadequate digital infrastructure, gaps in education and skills, and lagging policy frameworks, all of which contribute to limited access to AI tools and uneven capacity to develop AI solutions. The consequence of this divide is that many Africans could miss out on AI-driven innovations that drive economic growth and improve public services, thereby widening existing socio-economic gaps.

However, my exploration also highlights reasons for optimism. Initiatives like Deep Learning Indaba and Data Science Africa demonstrate a growing grassroots movement to democratise AI knowledge. Several African governments have begun crafting AI policies and investing in digital infrastructure, signalling political will to address the divide. International support and partnerships are increasingly aligning with Africa's digital needs, from connectivity projects to AI skill-building programs. The strategies recommended – from strengthening infrastructure and education to enacting inclusive policies – provide a roadmap for leveraging these positive trends and tackling the persistent challenges.

Looking ahead, there are rich avenues for future research. One important direction is to conduct empirical longitudinal studies on AI adoption in SSA and Africa as a whole – for instance, tracking the number of AI startups, research outputs, or AI-enabled services across different countries to measure progress in inclusivity quantitatively. Another area is impact assessment, which investigates how AI applications affect development outcomes in African contexts, such as analysing the impact of an AI agricultural advisory system on farmers' productivity or studying the employment effects of AI-driven automation in African industries. Furthermore, research can delve deeper into community perspectives on AI – using surveys or participatory methods to understand how various groups (rural communities, women, youth, and people with disabilities) perceive AI and what barriers they face in accessing it. Such studies would inform more tailored interventions.

In conclusion, bridging the AI divide in SSA is both a pressing challenge and a tremendous opportunity. By actively pursuing inclusive strategies, African nations can ensure that AI contributes to broad-based socio-economic development rather than exacerbating inequalities. The path to inclusivity will require collaboration among governments, academia, industry, and civil society and learning from local experiments and global best practices. If successful, Africa's AI journey could become a model for how emerging technologies can be harnessed in the service of inclusive growth, empowering communities that have historically been left behind in previous technological waves. The recommendations and insights in this paper aim to support that journey, emphasising that the AI revolution need not bypass Africa but can be shaped by and for Africa's diverse populace.